# Information-Theoretic Methods for Identifying Relationships among Climate Variables*

Kevin H. Knuth[1,2], Deniz Gencaga[1], William B. Rossow[3]
1. Department of Physics, University at Albany, Albany NY 12222
2. Department of Informatics, University at Albany, Albany NY 12222
3. Department of Electrical Engineering, The City College of New York, NY NY 10031

*Abstract*- **Information-theoretic quantities, such as entropy, are used to quantify the amount of information a given variable provides. Entropies can be used together to compute the mutual information, which quantifies the amount of information two variables share. However, accurately estimating these quantities from data is extremely challenging. We have developed a set of computational techniques that allow one to accurately compute marginal and joint entropies. These algorithms are probabilistic in nature and thus provide information on the uncertainty in our estimates, which enable us to establish statistical significance of our findings. We demonstrate these methods by identifying relations between cloud data from the International Satellite Cloud Climatology Project (ISCCP) and data from other sources, such as equatorial pacific sea surface temperatures (SST).**

## I. INTRODUCTION

The field of Earth-science is currently in the difficult stage of identifying relevant variables. How do we identify candidate variables or indices and ensure that they are maximally relevant to the phenomena we wish to describe or predict? To establish relevance between a variable and a phenomenon, one needs to demonstrate that the variable provides information about the phenomenon. To quantify this, we turn to information theory.

We have developed a suite of computational tools that allow researchers to calculate, from data, an information-theoretic quantity called entropy, which is critical to identifying relationships among climate variables. Our tools estimate entropy along with its associated error bars, the latter of which is critical for describing the degree of uncertainty in the estimates. This work builds upon our previous investigations of optimal binning techniques that we had developed for piecewise-constant, histogram-style models of the underlying density functions, and is more focused on continuous density models based on sums of Gaussians and Markov chain Monte Carlo sampling techniques.

## II. ENTROPY AND INFORMATION

We can characterize the behavior of a system $X$ by looking at the set of states the system visits as it evolves in time. If a state is visited rarely, we would be surprised to find the system there. We can express the expectation (or lack of expectation) to find the system in state $x$ in terms of the probability that it can be found in that state, $p(x)$, by

$$h(x) = \log \frac{1}{p(x)}. \quad (1)$$

This quantity is often called the surprise, since it is large for improbable events and small for probable ones. Averaging this quantity over all of the possible states of the system gives a measure of our expectation of the state of the system

$$H(X) = \sum_{x \in X} p(x) \log \frac{1}{p(x)}. \quad (2)$$

This quantity is called the Shannon Entropy, or entropy for short [1]. It can be thought of as a measure of the amount of information we possess about the system. It is usually expressed by rewriting the fraction above using the properties of the logarithm

$$H(X) = -\sum_{x \in X} p(x) \log p(x). \quad (3)$$

If the system states can be described with multiple parameters, the entropy can still be computed by averaging over all possible states (here it is shown for two subsystems $X$ and $Y$):

$$H(X,Y) = -\sum_{x \in X} \sum_{y \in Y} p(x,y) \log p(x,y). \quad (4)$$

This is called the joint entropy, since it describes the entropy of the states $X$ and $Y$, which jointly describe the system. One can often think of $X$ and $Y$ as representing subsystems of the original system. In this case, an important quantity is the difference of entropies,

$$MI(X,Y) = H(X) + H(Y) - H(X,Y). \quad (5)$$

This is called the Mutual Information (MI) since it describes the amount of information that is shared between the two subsystems. If you possess information about subsystem $X$, the mutual information describes how much information you also possess about $Y$. Thus mutual information is useful for quantifying the relevance of knowledge about one subsystem to knowledge about another subsystem, which is critical for quantifying predictability. For this reason, MI is critical in identifying relationships across climate variables, and in identifying and selecting a set of relevant variables that aid in the prediction of another climate variable. If two climate variables *X* and *Y* are independent, then

$$H(X,Y) = H(X) + H(Y), \quad (6)$$

and the mutual information (5) is zero.

* Supported by NASA AIST-QRS-07-0001

## III. ESTIMATING ENTROPIES

### A. Density Models

Estimation of these information-theoretic measures from data is, at first glance, deceivingly straightforward. From the data one estimates the probability density from which the data were sampled. Then using this probability density, one can use the formulae above to estimate the various necessary entropies. Most researchers focus on the biases that are introduced in the calculations (which are virtually impossible to estimate without prior information) and neglect computing error bars on the entropy estimates. This results in estimates with unknown statistical significance. To accurately estimate the associated uncertainties, one must propagate the uncertainties in the probability density estimation through the equations for the entropy and mutual information. Consequently, this procedure relies heavily of estimating probability density functions and keeping track of our uncertainties.

Our early efforts focused on piecewise-constant (histogram-like) density models [2, 3]; the reason being that these models are extremely fast computationally. While this was found to be true, we also found that the constituent uniform densities led to small biases in the entropy estimates, which are disastrous in higher dimensions. This is important since the mutual information depends on the joint entropy, which depends on a two-dimensional probability density, and quantities like the transfer entropy [4] depends on higher dimensions still.

Since the main problem with piecewise-constant densities was the discrete nature of the bins, we have adopted a Mixture of Gaussians model (MoG) where the probability density is modeled as a sum of Gaussian densities. Given a set of data, we fit a density function to it based on a sum of Gaussians of varying means and standard deviations.

### B. Entropy Estimates from Density Estimates

With this density function in hand, we can then compute the entropy by a standard numeric integration with the integral version of equation (3) or (4) above. However, this alone neither provides us with an optimal estimation of the entropy nor an estimate of our uncertainty. To obtain these, we need a set of density functions sampled from the posterior probability of density functions. This turns out to be extremely easy in the case of piecewise-constant density models since the model parameters can be sampled directly from a Dirichlet distribution and can be accomplished directly without resorting to more sophisticated sampling techniques.

When working with the MoG models, we rely on the new Markov chain Monte Carlo (MCMC) algorithm called Nested Sampling [5]. By integrating this MCMC algorithm with some basic sampling code, which selects representative MoG models from the MCMC samples according to the posterior probability, we obtain a set of probable density function models.

Once a set of probability density models are obtained from the data, we can obtain a set of entropies. Computing the entropy for each model and weighting that particular entropy calculation by the probability of the model enables us to compute the mean entropy and the standard deviation. Our experiments with one-dimensional Gaussian-distributed data verifies that this method enables us to accurately estimate entropies from data within the error bars 67% of the time in the case of piecewise-constant models and 68% of the time in the case of MoG models. As described earlier, in two-dimensional data sets, the piecewise-constant models result in biases in the entropy estimation that do not appear when using the MoG models.

## IV. MUTUAL INFORMATION

We have performed experiments with Gaussian distributions and demonstrated that we obtain mutual information estimates within the error bars for non-correlated data sets. For correlated data sets the estimates are within error bars only 24% of the time, which suggests that we are underestimating the uncertainties. One example is the case where the correlated data had a true mutual information of 0.1438, which we estimated to be 0.1361 ± 0.0050. This over-confidence is most likely due to the fact that the MI is a positive quantity (bounded from below); although it could be that our 2-D integration scheme is not sufficiently precise. These issues continue to be investigated.

We now demonstrate the utility of these techniques by examining the MI between interesting climate variables. We consider the percent cloud cover as subsystem *X*. These data were obtained from the International Satellite Cloud Climatology Project (ISCCP) climate summary product C2 [5, 6], and consist of monthly averages of percent cloud cover resulting in a time-series of 198 months of 6596 equal-area pixels each with side length of 280 km. The percent cloud cover data at each pixel can be thought of as a time series of measurements of subsystem *X*: $X_1, X_2, \ldots, X_{6596}$. The other data set was chosen to be the Cold Tongue Index (CTI), which describes the sea surface temperature anomalies in the eastern equatorial Pacific Ocean (6N-6S, 180-90W) [7]. These anomalies are indicative of the El Niño Southern Oscillation (ENSO) [8, 9]. These data, which consist of the set of 198 monthly values of CTI, constitute the subsystem *Y*. The data *X* and *Y* were chosen to correspond in time.

The MI was computed between the cloud cover at pixel 1

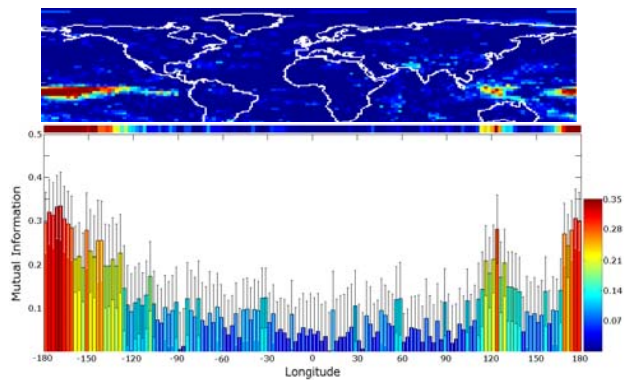

**Figure 1**. The top shows a Mutual Information (MI) map of Cloud Cover vs. Eastern Equatorial Pacific CTI, which indexes ENSO, generated using a piecewise-constant density model. The bottom shows the MI between CTI and Cloud Cover along the equator. The error bars, enable us to identify statistically significant dependencies.

($X_1$) and the CTI ($Y$), and pixel 2 ($X_2$) and the CTI ($Y$), and so on by using (12). This enables us to generate a global map of 6596 mutual information calculations, which indicates the relationship between the Cold Tongue Index (CTI) and percent cloud cover across the globe.

Fig. 1 shows the mutual information as estimated using piecewise-constant density models. Below the map is a bar graph that indicates the mutual information estimates for sites along the equator. The error bars indicate the uncertainties in our estimates, which can be used to identify statistically significant regions and eliminate those that are not significant. Fig. 2 shows the mutual information map with only the regions with that have two-sigma significance and greater. From this map, one can see that the cloud cover affected by the sea surface temperature (SST) variations lies mainly in the equatorial Pacific, along with an isolated area in Indonesia. The highlighted areas in the Indian longitudes are known artifacts of satellite coverage.

Fig. 3 shows the same map, but estimated using the Mixture of Gaussian model. While it is known that the MoG model will produce more accurate mutual information estimates, we are confident that the uncertainties are underestimated. For this reason, the two-sigma cut-off still permits a large amount of noise in the map. However, despite this noise, more structure is revealed. For instance, note the effect on the inter-tropical convection zones across South America and Africa. It is known that seasonality also affects the cloud cover in these regions [10], so it is unclear as to what degree these high mutual information values are due to the fact that both ENSO and the percent cloud cover are functions of seasonality, or whether ENSO is modulating the percent cloud cover in these areas. To establish this, we would need to move on to higher-order informations, such as transfer entropy [4].

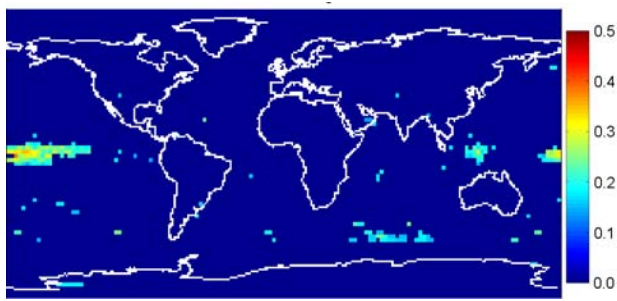

**Figure 2**. A thresholded version of the map in Fig. 1 of the effect of ENSO on cloud cover including only pixels that exhibit statistical significance of at least two standard deviations.

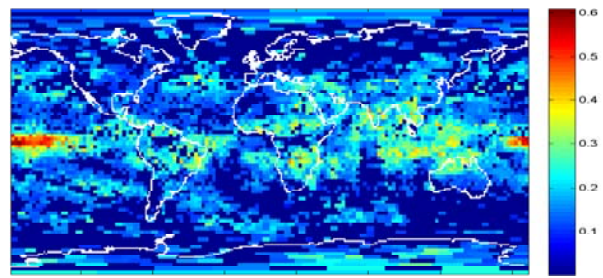

**Figure 3**. A Mutual Information map of the ENSO SST and ISCCP cloud cover data using the Mixture of Gaussians density model. At the two standard deviation level of significance, more regions are seen to be related, such as the inter-tropical convection zones.